\documentclass[aps,prb,showpacs,twocolumn]{revtex4-1}

\usepackage{amsmath,amssymb,stmaryrd} 

\usepackage{ifpdf} 
\ifpdf 
  \usepackage[pdftex,breaklinks=true]{hyperref}
  \usepackage{graphicx} 
\else  
  \usepackage[hypertex,breaklinks=true]{hyperref}
  \usepackage{graphicx} 
\fi

\newcommand{\Tr}{\mathop{\mathrm{Tr}}\nolimits}
\renewcommand{\vec}[1]{\boldsymbol{#1}}
\renewcommand{\Re}{\mathop{\mathrm{Re}}\nolimits} 
\renewcommand{\Im}{\mathop{\mathrm{Im}}\nolimits} 

\begin{document}

\title{Quantum thermodynamics of nanoscale steady states far from equilibrium}
 
\author{Nobuhiko Taniguchi}
\email{taniguchi.n.gf@u.tsukuba.ac.jp}
\affiliation{Physics Division, Faculty of Pure and Applied Sciences,
  University of Tsukuba, Tennodai Tsukuba 305-8571, Japan}

\date{\today}

\begin{abstract}
  We develop an exact quantum thermodynamic description for a
  noninteracting nanoscale steady state that couples strongly with
  multiple reservoirs.  It is demonstrated that there exists a
  steady-state extension of the thermodynamic function that correctly
  accounts for the multiterminal Landauer-B\"{u}ttiker formula of
  quantum transport of charge, energy or heat, via the nonequilibrium
  thermodynamic relations.  Its explicit form is obtained for a single
  bosonic or fermionic level in the wide-band limit, and corresponding
  thermodynamic forces (affinities) are identified.  Nonlinear
  generalization of the Onsager reciprocity relations are derived.  We
  suggest that the steady-state thermodynamic function is also capable
  of characterizing the heat current fluctuations of the critical
  transport where the thermal fluctuations dominate. It is also
  pointed out that the suggested nonequilibrium steady-state
  thermodynamic relations seemingly persist for a spin-degenerate single level
  with local interaction.
\end{abstract}



\maketitle



\section{Introduction}

Constructing a thermodynamic theory that applies consistently to
nonequilibrium steady states has long been a theoretical challenge in
many fields of science, not only in physics but also in chemistry or
biology.  Steady states differ from equilibrium states by being driven
by the external environments (reservoirs) and accommodating finite
flows that induce entropy production.  Formulating thermodynamics for
such irreversible systems is notoriously difficult; successes have
been mainly achieved within the linear response theory, where various
transport coefficients can be related to fluctuations in
equilibrium~\cite{Onsager53,KuboBook85}.  Beyond the linear-response regime,
a possible thermodynamic formulation has been anticipated for the steady
states~\cite{Oono98,Sasa06,Komatsu08,Saito11c,Hyldgaard12}. Yet the
theory has been largely unexplored so far, partly because basic
concepts such as temperature and entropy get elusive and questioned
when treating an open, irreversible system.

\begin{figure}
  \centering
  \includegraphics[width=0.8\linewidth]{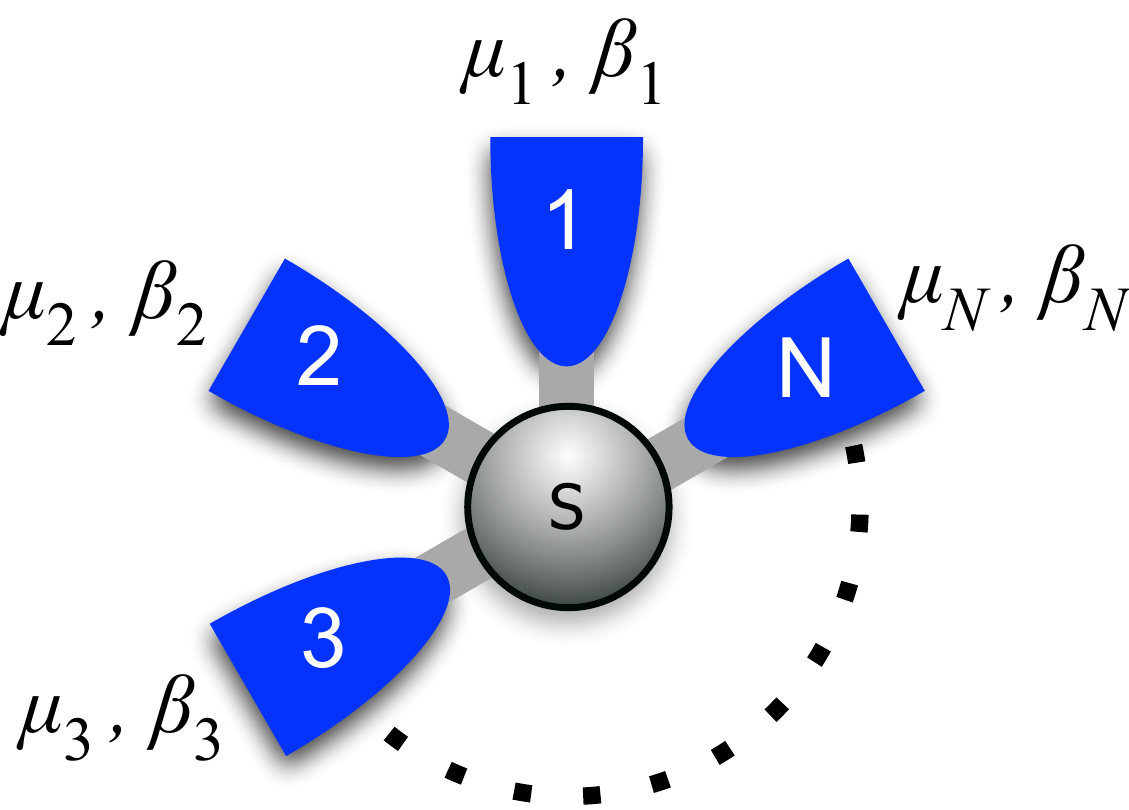}
  \caption{An open system (S) connected with multiple reservoirs with
    different inverse temperatures $\beta_{a}$ and chemical potentials
    $\mu_{a}$ for $a = 1, \ldots, N$. }
\label{fig:multiterminal}
\end{figure}

In recent years, it has been recognized that thermodynamic laws are
consistent with quantum properties of open nanoscale systems, 
typically connected with multiple reservoirs with different
chemical potentials and temperatures
(Fig.~\ref{fig:multiterminal}).  The emergence of thermodynamics 
is somewhat unexpected; the situation is opposite
to the conventional thermodynamic limit, as it involves only a
few particles, even a single one.
The statistical ensemble average is replaced by quantum averaging, and
the applicability of thermodynamics results from its quantum nature.
Nanoscale systems provide a rare and novel opportunity to study the
steady-state thermodynamics, without relying on any statistical ensemble
hypothesis~\cite{Anders17}.

As a realization of nonequilibrium systems, a steady-state nanoscale
system is notable because of its strong coupling with the
reservoirs. One cannot rely on local equilibrium hypothesis to
characterize a nano-system. Any temperature or chemical potential
cannot be assigned to it \emph{a priori}, except for a system coupled with a
single reservoir.  Strongly coupled reservoirs make
effective dynamics non-Markovian with memory effect, and the
Lindblad form of the master equation invalidated.
The nonequilibrium density matrix is represented not by the
standard Gibbs ensemble but by a generalized one (the
MacLennan-Zubarev type)~\cite{ZubarevBook96,Tasaki06,Fujii07,Ness13}.
A consistent thermodynamic framework is nontrivial even for
noninteracting transport. It has been striven for by many
approaches~\cite{Ludovico14,Esposito15,Topp15,Bruch16,Stafford17}, but
it has not been fully disclosed so far.

In this paper, we develop a thermodynamic description of the steady
state at nanoscale, entirely based on quantum mechanics.  A
steady-state extension of the Massieu-Planck function
$\Phi_{\text{ss}}$, which is determined by normalization of the
reduced density matrix, is singled out as a nonequilibrium
thermodynamic function.
One salient feature of this quantum construction is that the resulting
thermodynamic function $\Phi_{\text{ss}}$ unavoidably becomes
stationary in time because the steady-state density matrix is
time-independent.  It contrasts sharply with a naive expectation that
a steady-state thermodynamic function might increase in time because
it may include an increasing entropic contribution.  Notwithstanding
we will demonstrate that the function $\Phi_{\text{ss}}$ is viable to
describe the steady-state properties far from equilibrium.
To fully characterize the steady state, we also need to identify a
correct set of parameters that control the steady state. A
consistent choice of them includes local (inverse) temperature and
chemical potential, $(\bar{\beta}, \overline{\beta\mu})$, as well as
various affinities $(A^{N}_{a}, A^{E}_{a})$ that are thermodynamic
forces to drive the system out of equilibrium [see
Eqs.~\eqref{def:affinities} for their definitions]. With these
parameters, the significance of $\Phi_{\text{ss}}$ is
compactly represented in the differential form:
\begin{align}
& d\Phi_{\text{ss}} = \bar{N} d (\overline{\beta \mu}) - \bar{E} d \bar{\beta}
+ \frac{\hbar}{2\gamma}\sum_{a} \left( I_{a} \, d A^{N}_{a} + J_{a} \, d
  A^{E}_{a} \right),
\label{eq:dPhi-ss}
\end{align}
where $\bar{N}$ or $\bar{E}$ is the average occupancy or energy of the
system, while $I_{a}$ or $J_{a}$ is a nonlinear inflow of particle or
energy from the reservoir $a$. The constant $\gamma$ is the total
relaxation rate of the system.
The relation \eqref{eq:dPhi-ss} serves as a nonequilibrium extension
of the thermodynamic relation of the Massieu-Planck function (see
Appendix \ref{sec:MP-thermodynamics}).
Being stationary in time, the function $\Phi_{\text{ss}}$ does not
refer to the internal entropy. Yet $\Phi_{\text{ss}}$ 
characterizes the entropy production rate by
\begin{align}
& \frac{dS}{dt} = \frac{2\gamma}{\hbar} \sum_{a} \left( A^{E}_{a}
  \frac{\partial \Phi_{\text{ss}}}{\partial A^{E}_{a}} + A^{N}_{a}
  \frac{\partial \Phi_{\text{ss}}}{\partial A^{N}_{a}}
  \right). 
\label{eq:entropy-rate-by-Phi}
\end{align}
These formulas~\eqref{eq:dPhi-ss}--\eqref{eq:entropy-rate-by-Phi} will
be proved to be exact for a noninteracting single bosonic or fermionic
level that couples linearly with multiple reservoirs.  Moreover we
argue the above thermodynamic structure
Eqs.~\eqref{eq:dPhi-ss}--\eqref{eq:entropy-rate-by-Phi}, found in a
noninteracting steady state, persists even in a steady state of the
model with local interaction, namely, the single-impurity Anderson
model.

\section{Model and known results}

The total Hamiltonian consists of $H = H_{S} + H_{R} + H_{SR}$, whose
terms represent a nanoscale system (``quantum
dot''), multiple reservoirs with different inverse temperatures
$\beta_{a}$ and chemical potentials $\mu_{a}$ (for $a=1, \ldots, N$),
and the linear coupling between the system and the reservoirs (see
Fig.~\ref{fig:multiterminal}).
They are 
\begin{align}
& H_{S} = \epsilon_{d}\, d^{\dagger} d, \\
& H_{R} = \sum_{a} \epsilon_{a\vec{k}}\,  c_{a \vec{k}}^{\dagger}
  c_{a \vec{k}}, \\
& H_{SR} = \sum_{a,\sigma} \left( V_{d a}\, d^{\dagger}
  c_{a\vec{k}} + V_{ad}\, c^{\dagger}_{a\vec{k}}
  d \right),
\end{align}
where $d^{\dagger}$ creates a particle with energy $\epsilon_{d}$ at
the system and $c_{a\vec{k}}^{\dagger}$, with energy
$\epsilon_{a \vec{k}}$ at the reservoir $a$.  Particles can be bosonic
or fermionic.  We present the results for both cases simultaneously
with composite signs (with the upper for bosonic; the lower for
fermionic).

The presence of the reservoir makes the nanoscale system dissipative,
inducing a finite resonant width $\gamma_{a} = \pi |V_{da}|^{2} \rho_{a}$
due to the reservoir $a$ (with its density of states $\rho_{a}$). 
Quantum transport across a noninteracting system can be solved exactly
by several approaches, such as the scattering method, the
equation-of-motion method, or the nonequilibrium Green function
method.  When we take the wide-band approximation, the inflow of
particle $I_{a}$ or of energy $J_{a}$ from the reservoir $a$ is given
by the Landauer-B\"{u}ttier formula~\cite{Landauer57,Sivan86,Meir92}:
\begin{align}
& I_{a} = \frac{2}{\hbar} \sum_{b}
  \frac{\gamma_{a} \gamma_{b}}{\gamma}
  \int d\varepsilon\, \rho(\varepsilon) \, \left[ f_{a}(\varepsilon) -
  f_{b}(\varepsilon) \right],
\label{eq:LB1} \\
& J_{a} = \frac{2}{\hbar} \sum_{b}
  \frac{\gamma_{a} \gamma_{b}}{\gamma}
  \int d\varepsilon\, \rho(\varepsilon) \varepsilon \, \left[
  f_{a}(\varepsilon) - f_{b}(\varepsilon) \right].
\label{eq:LB2}
\end{align}
Here $\rho(\varepsilon)$ refers to the spectral function of the system, 
\begin{align}
& \rho(\varepsilon) = \frac{\gamma/\pi}{(\varepsilon -
  \epsilon_{d})^{2} + \gamma^{2}}; \quad \gamma = \sum_{a} \gamma_{a},
\label{def:spectral-fn}
\end{align}
and
$f_{a}(\varepsilon) = [e^{\beta_{a} (\varepsilon - \mu_{a}) } \mp
1]^{-1}$ is the distribution function of the reservoir $a$.  These
currents are usually expressed by the transmission
$\mathcal{T}_{ab}(\varepsilon) = 4\pi \gamma_{a} \gamma_{b}
\rho(\varepsilon)/\gamma$ between the reservoirs $a$ and $b$, but we
prefer writing them in the above form.
One can evaluate them analytically in terms of the digamma function
[See Eqs.~(\ref{eq:I-analytical})--(\ref{eq:J-analytical})].  Heat
current flowing from the reservoir $a$ is defined by
$J^{Q}_{a} = J_{a} - \mu_{a} I_{a}$.
The average number $\bar{N}$ and energy $\bar{E}$ of
the system are given by 
\begin{align}
& \bar{N} 
= \int d\varepsilon \, \sum_{a} \frac{\gamma_{a}}{\gamma} \rho(\varepsilon)
 f_{a}(\varepsilon), 
  \label{eq:Nbar}\\
& \bar{E} 
= \int d\varepsilon \, \sum_{a} \frac{\gamma_{a}}{\gamma}
  \rho(\varepsilon)\, \varepsilon\, 
 f_{a}(\varepsilon).
\label{eq:Ebar}
\end{align}

We will show below that the steady-state thermodynamic function
$\Phi_{\text{ss}}$, which is constructed quantum mechanically,
correctly produces quantum transport
equations~\eqref{eq:LB1}--\eqref{eq:LB2} as well as local 
quantities~\eqref{eq:Nbar}--\eqref{eq:Ebar} via the differential
relation Eq.~\eqref{eq:dPhi-ss}.

\section{The Massieu-Planck function}

Analogous to an equilibrium system, our basic assumption is that the
partition function which normalizes the density matrix bridges between
a microscopic model and its thermodynamics.  We suppose its
steady-state extension is provided instead by normalizing the reduced
density matrix $\hat{\varrho}_{\text{red}}$ of the relevant system.
In treating the steady state, we find it
advantageous to use the Massieu-Planck function~\cite{CallenBook85},
which is defined by the logarithm of the (effective) partition
function. 

\subsection{Single-reservoir Massieu-Plank function}

As for an open system that connects with a single reservoir with
$\beta$ and $\mu$, the effective thermodynamics has long
been investigated~\cite{Feynman63,Schotte75,%
Caldeira83,WeissBook08,Hanggi08,Adamietz14}.
By recasting it, the single-reservoir
Massieu-Planck function is found to be (see Appendix \ref{sec:MP1-thermodynamics})
\begin{align}
& \Phi_{1} (\beta,\beta\mu) =  \mp \int^{\infty}_{-\infty} d\varepsilon \, \rho(\varepsilon) \ln
  \left[ 1 \mp e^{-\beta(\varepsilon - \mu)}\right].
\label{def:Phi1}
\end{align}
The energy integration actually diverges in the wide-band limit, so
some regularization is needed. In Appendix \ref{sec:MP1-analytical}, we show the explicit analytical
form of $\Phi_{1} (\beta, \beta\mu)$ with regularization, and examine
its various thermodynamic properties that are
independent of regularization.
The physics of $\Phi_{1}$ is transparent; the level of the
open nano-system acquires finite broadening due to coupling with the
reservoir.
We make a point of regarding $\Phi_{1}$ as a function of $\beta$ and
$\beta\mu$, as they are parameters dual to particle number and energy.
We stress that they are originally external parameters specified by
the reservoir.  The implication of $\beta$ and $\mu$ as thermodynamic
parameters is somewhat blurred because the reduced density matrix
$\varrho_{\text{red}}$ is no longer represented by the standard Gibbs
ensemble.

\subsection{Steady-state Massieu-Planck function}

One can calculate the steady-state Massieu-Planck function
$\Phi_{\text{ss}}$ that couples with multiple reservoirs by
normalizing the reduced density matrix $\varrho_{\text{red}}$.  As we
work on noninteracting systems, the calculation can drastically be
simplified by utilizing a Gaussian nature of
$\hat{\varrho}_{\text{red}}$, in light of the Zubarev's relevant
distributions and nonequilibrium statistical
operators~\cite{ZubarevBook96} (see
also~\cite{Chung01,Peschel03,Dhar12}).
Asking $\bar{N} = \langle d^{\dagger} d \rangle$ to reproduce
Eq.~\eqref{eq:Nbar}, we deduce that $\hat{\varrho}_{\text{red}}$ may
well be represented in terms of relevant field operators
$\psi_{a}(\varepsilon)$ and $\psi_{a}^{\dagger}(\varepsilon)$, satisfying
\begin{align}
\label{eq:relevant-field-comm}
& \left[ \psi_{a}(\varepsilon), \psi_{a'}^{\dagger}(\varepsilon') \right]_{\mp}
= \delta_{aa'}\, \delta(\varepsilon - \varepsilon'), \\
& \left\langle \psi_{a'}^{\dagger} (\varepsilon')
  \psi_{a}(\varepsilon)\right\rangle
= \delta_{aa'}\, \delta(\varepsilon - \varepsilon')\,
  f_{a}(\varepsilon), \\
& d = \sum_{a} \int d\varepsilon \sqrt{\frac{\gamma_{a}}{\gamma}
  \rho(\varepsilon)} \,   \psi_{a}(\varepsilon).
\label{eq:d-in-psi}
\end{align}
Then the function $\Phi_{\text{ss}}$ normalizes
$\hat{\varrho}_{\text{red}}$ as
\begin{align}
  & \hat{\varrho}_{\text{red}} = e^{-\Phi_{\text{ss}}
    -\sum_{a,\varepsilon} \beta_{a} (\varepsilon - \mu_{a}) \psi_{a}^{\dagger}(\varepsilon)
    \psi_{a}(\varepsilon)},
    \label{eq:rho-red}
\end{align}
where
$\sum_{a,\varepsilon}(\cdots) = \sum_{a}\int d\varepsilon
(\gamma_{a}/\gamma) \rho(\varepsilon) (\cdots)$ is the
summation/integral over the energy shell and the reservoirs. 
Determining $\Phi_{\text{ss}}$ by imposing
$\Tr \hat{\rho}_{\text{red}} = 1$ is equivalent to evaluating the
functional determinant. A quick, symbolic way to evaluate it is
\begin{align}
  & \exp \left( \Phi_{\text{ss}} \right)
    = \prod_{a,\varepsilon} \left[ 1 \mp e^{-\beta_{a} (\varepsilon -\mu_{a})}
      \right]^{\mp 1}
\label{eq:rho-red-det}
,\\ & \quad
= e^{\mp \sum_{a,\varepsilon} \ln \left[ 1 \mp e^{-\beta_{a}
      (\varepsilon -\mu_{a})}  \right]}.
\end{align}
It expresses the steady-state Massieu-Planck function,
\begin{align}
& \Phi_{\text{ss}} = \sum_{a} \frac{\gamma_{a}}{\gamma}
  \Phi_{1} (\beta_{a}, \beta_{a}\mu_{a}),
  \label{def:Phi-ss}
\end{align}
as a superposition of the single-reservoir contribution
$\Phi_{1}(\beta_{a},\beta_{a}\mu_{a})$. Hence $\Phi_{\text{ss}}$ can be evaluated
analytically.  The manipulation
of Eq.~\eqref{eq:rho-red-det} is due to observing that
$(\gamma_{a}/\gamma) \rho(\varepsilon)$ acts as (fractional)
degeneracies satisfying $\sum_{a,\varepsilon} 1 = 1$; such analytical
continuation is validated because it correctly reproduces the
single-reservoir result~\eqref{def:Phi1}.

The relevant field operator $\psi_{a}^{\dagger} (\varepsilon)$ in
Eqs.~(\ref{eq:relevant-field-comm}--\ref{eq:rho-red}) has a clear
physical meaning.  One can construct the steady-state density matrix
$\hat{\varrho}_{\text{ss}}$ of the total system (the system plus the
reservoirs)~\cite{Hershfield93,Frohlich03,Tasaki06,Oguri07,Ness13,Ness14b},
\begin{align}
& \hat{\varrho}_{\text{ss}} 
  \propto \exp \left[ -\sum_{a}\sum_{\vec{k}}\beta_{a}
  (\epsilon_{a\vec{k}} - \mu_{a}) \psi_{a\vec{k}}^{\dagger} \psi_{a\vec{k}}
  \right], 
\label{eq:rho-ss}
\end{align}
where
$\psi_{a\vec{k}}^{\dagger} = \Omega\, c_{a\vec{k}}^{\dagger}
\Omega^{\dagger}$ is a scattering-state field of the reservoir $a$
that is defined by the M\o ller operator $\Omega$.  The field
$\psi^{\dagger}_{a\vec{k}}$ becomes a coherent superposition of fields
$c^{\dagger}_{a\vec{k}}$ and $d^{\dagger}$. Accordingly, field $d$ is
solved to be a superposition of the scattering fields involving all
the reservoir fields, as in Eq.~\eqref{eq:d-in-psi}. It accounts for
quantum coherence between the system and the reservoirs.
The average density is
$\langle \psi_{a\vec{k}}^{\dagger} \psi_{a\vec{k}}\rangle =
f_{a}(\epsilon_{a\vec{k}})$, and the canonical (anti-)commutation
relation is preserved.  The relevant field $\psi_{a}(\varepsilon)$ is
nothing but an energy representation of the scattering-state field
$\psi_{a\vec{k}}$.

We cannot emphasize too much a novel and peculiar nature of
Eq.~\eqref{def:Phi-ss}.  
Although such a superposition is a common trait of quantum mechanics,
Eq.~\eqref{def:Phi-ss} tells that 
the function $\Phi_{\text{ss}}$ that describes the irreversible steady state 
(with the increasing entropy) is unchanged in time and given
by a superposition of $\Phi_{1}$'s of the single
reservoirs, each of which refers to the entropy-preserving, reversible
system.  One may notice such a trait of superposition in the expression of
average number $\bar{N}$ or energy $\bar{E}$ [Eq.~\eqref{eq:Nbar} or
\eqref{eq:Ebar}], but it is far from obvious that one can use
$\Phi_{\text{ss}}$ to describe quantum transport $I_{a}$
and $J_{a}$.  To fully disclose the steady-state thermodynamics, 
one needs to find what are relevant controlling parameters for it.

\subsection{Affinities}

Finding the correct set of appropriate controlling parameters
arbitrary away from equilibrium is quite nontrivial, but it is
imperative to establish the steady-state thermodynamic relations.  The
function $\Phi_{\text{ss}}$ of Eq.~\eqref{def:Phi-ss} depends on $2N$
independent external parameters
$\{\beta_{a}, \mu_{a}\}_{a=1, \ldots, N}$ specified by the reservoirs.
Among them, we expect that two \emph{local} parameters (temperature
and chemical potential) regulate the average particle number and
energy [Eqs.~\eqref{eq:Nbar}--\eqref{eq:Ebar}], while all other $2N-2$
parameters (the difference of temperatures and/or chemical potentials)
drive the system out of equilibrium and cause irreversible
processes. The latter parameters are called thermodynamic forces or
affinities.  Among them, we will identify the relevant parameters
$(\bar{\beta}, \overline{\beta\mu}, \{A^{N}_{a}, A^{E}_{a}\})$ defined
in Eqs.~\eqref{def:affinities} below, which describe the quantum
transport as well as thermodynamic properties. This constitutes our
main result, with the steady-state thermodynamic
function~\eqref{def:Phi-ss}. 

One can identify affinities and their associated currents by examining
the internal entropy production rate~\cite{CallenBook85}.  In
the system we consider, it is balanced with the entropy inflow,
so that we find
\begin{align}
& \frac{dS}{dt} = 
-\sum_{a}  \beta_{a} \left( J_{a} - \mu_{a} I_{a} \right).
  \label{eq:entropy-rate-LB}
\end{align}
Its positivity follows because the distribution function
$f_{a}(\varepsilon)$ is a decreasing function regarding
$\beta_{a} (\varepsilon
-\mu_{a})$~\cite{Frohlich03,Tasaki06,Nenciu07,Topp15}.
The form of Eq.~\eqref{eq:entropy-rate-LB} tells us to introduce two
types of affinities associated with each reservoir: chemical
affinities $A^{N}_{a}$ to generate particle currents, which is a
deviation of $\beta \mu$, and thermal affinities $A^{E}_{a}$ to
generate energy currents, which is a deviation of $-\beta$. Those
deviations must be defined from some reference values, $\bar{\beta}$
and $\overline{\beta\mu}$, which in turn regulate $\bar{N}$ and
$\bar{E}$. 
We choose to introduce affinities for conserved currents of particle
and energy rather than heat currents.  The
conservation laws are fulfilled by the condition
$\sum_{a} \gamma_{a} A^{E}_{a} = \sum_{a} \gamma_{a} A^{N}_{a} = 0$. 
Hence a pair of $A^{N,E}_{a}$ are redundant (see Appendix
\ref{sec:some-detail}
for an explicit construction).
All things considered, we come to make the following choice of local
quantities $\bar{\beta}$ and $\overline{\beta\mu}$, and
affinities:
\begin{subequations}
\label{def:affinities}
\begin{align}
& A^{N}_{a} = \beta_{a} \mu_{a} - \overline{\beta \mu}; 
&& \overline{\beta\mu} = \sum_{a} \frac{\gamma_{a}}{\gamma}\, \beta_{a}
  \mu_{a}, \\
&A^{E}_{a} = -\beta_{a} + \bar{\beta}; 
&& \bar{\beta} = \sum_{a} \frac{\gamma_{a}}{\gamma}\, \beta_{a}.
\end{align}
\end{subequations}

Viewing $\Phi_{\text{ss}}$ of Eq.~\eqref{def:Phi-ss} as a function of
the above set of parameters
$(\bar{\beta}, \overline{\beta\mu}, \{A^{N}_{a}, A^{E}_{a}\})$, it is
now straightforward to find (see Appendix \ref{sec:some-detail} for
details)
\begin{align}
&\bar{N} = \frac{\partial \Phi_{\text{ss}}}{\partial (\overline{\beta\mu})}; 
\quad \qquad \bar{E} = -\frac{\partial \Phi_{\text{ss}}}{\partial
  \bar{\beta}}, \label{eq:NE-by-Phi}
\\
& I_{a} = \frac{2\gamma}{\hbar} \frac{\partial \Phi_{\text{ss}}}{\partial
  A^{N}_{a}}; \qquad
J_{a} = \frac{2\gamma}{\hbar} \frac{\partial \Phi_{\text{ss}}}{\partial
  A^{E}_{a}}. 
\label{eq:currents-by-Phi}
\end{align}
The expression Eq.~\eqref{eq:entropy-rate-by-Phi} of the entropy production rate $dS/dt$ follows from
Eqs.~\eqref{eq:entropy-rate-LB}--\eqref{eq:currents-by-Phi}. 

\section{Discussion}

\subsection{Local temperature}

Our definition of local temperature $\bar{\beta}$ and chemical
potential $\bar{\mu} = \overline{\beta\mu}/\bar{\beta}$ is motivated
by the theoretical consistency of the thermodynamic formulation.  
Alternatively, one may probe local quantities by
measurements such as the scanning thermal technique~\cite{Majumdar99}.
Those \emph{probed} quantities, $\beta_{p}$ and $\mu_{p}$, are determined
by the no-flow condition of charge and energy when attaching the probe
reservoir~\cite{Meair14,Stafford16,Shastry16}:
\begin{align}
& \int d\varepsilon\, \rho(\varepsilon) \, \varepsilon^{k} \left[
  \bar{f} (\varepsilon) -  f_{p}(\varepsilon)  \right] = 0 \quad
  \text{(for $k=0,1$)}, 
\end{align}
where  $\bar{f}(\varepsilon)=\sum_{a} (\gamma_{a}/\gamma)
f_{a}(\varepsilon)$ is the effective distribution of the
reservoirs. 
In a general nonlinear setting far from equilibrium, parameters $\beta_{p}$ and
$\mu_{p}$ may differ from $\bar{\beta}$ and $\bar{\mu}$.  However,
as for the linear deviation, the probed quantities $\beta_{p}$ and
$\mu_{p}$ agree with $\bar{\beta}$ and $\bar{\mu}$, because the
effective distribution $\bar{f}(\varepsilon)$ can be expanded as
$\bar{f}(\varepsilon) \approx f_{p}(\varepsilon) + f'_{p}(\varepsilon)
[(\bar{\beta} - \beta_{p})\varepsilon - (\overline{\beta\mu} -
\beta_{p} \mu_{p})]/\beta_{p}$.
We also note that the scale $\bar{\mu}$ has played
an important role of characterizing nonlinear electronic transport in the Kondo
regime through an interacting dot~\cite{Taniguchi14}.

Explicit forms of local quantities and affinities in
Eqs.~\eqref{def:affinities} are outcomes of the wide-band
approximation, which is well justified for quantum coherent transport
through a nanostructure.  If $\gamma_{a}$ were to acquire substantial
energy dependence, one could nonetheless construct $\Phi_{\text{ss}}$
by generalizing Eq.~\eqref{def:Phi-ss} to take an energy-dependent
superposition for each energy shell.  However, it is quite
nontrivial in this situation how to identify appropriate 
controlling parameters that enable us to construct the
thermodynamic description.

\subsection{Maxwell relations and nonlinear generalization of the
  Onsager relations}

The existence of the function $\Phi_{\text{ss}}$ that satisfies the
differential form Eq.~\eqref{eq:dPhi-ss} has important consequences
for the steady-state thermodynamic structure. One can derive various
steady-state extensions of Maxwell relations by using the symmetry of
second derivatives.  
For instance, we see the
$\overline{\beta\mu}$ dependence of the current can be obtained by the
chemical affinity dependence of the occupancy, as in
\begin{align}
& \frac{\partial^{2} \Phi_{\text{ss}}}{\partial A^{N}_{a} \partial
  (\overline{\beta \mu})}
= \frac{\partial \bar{N}}{\partial A^{N}_{a}}
= \frac{\hbar}{2\gamma}\cdot \frac{\partial I_{a}}{\partial
  (\overline{\beta\mu})}.
\end{align}
Many other relations are derived similarly. One can furthermore make a
nonlinear generalization of the Onsager reciprocity relations by
the symmetry
$\partial^{2} \Phi_{\text{ss}}/\partial A^{E}_{a} \partial A^{N}_{a}$:
\begin{align}
& \frac{2\gamma}{\hbar}\cdot \frac{\partial \Phi_{\text{ss}}}{\partial
  A^{E}_{a} \partial  A^{N}_{a}}
= \frac{\partial I_{a}}{\partial A^{E}_{a}} = \frac{\partial
  J_{a}}{\partial A^{N}_{a}},  
\end{align}
which is valid for nonlinear responses. In the linear-response limit
(or the zero-affinity limit), the above gives the usual Onsager
reciprocal relations between the cross coefficients.

\subsection{Implication in the interacting system}

We have demonstrated that the function
$\Phi_{\text{ss}}$ characterizes the steady state, based on a
noninteracting transport model through a single level.
Notwithstanding the validity of the 
thermodynamic structure
Eqs.~(\ref{eq:dPhi-ss},\ref{eq:entropy-rate-by-Phi}) seems to go beyond
noninteracting systems to include a steady-state with local
interaction. Let us consider the spin-degenerate fermionic single
level with local interaction connecting with the multiple reservoirs,
namely, the nonequilibrium single-impurity Anderson model.
For that system, we can still derive the Landauer-B\"{u}ttiker type
formulas~\eqref{eq:LB1}--\eqref{eq:LB2} by help of nonequilibrium
Green functions~\cite{Meir92,Lebanon01,Sun01,Tolea09,Taniguchi14},
where many-body effect is encapsulated only in the spectral
function $\rho(\varepsilon) = - \Im G^{R}(\varepsilon)/\pi$.
Moreover, in the wide-band limit, the current conservation of charge
and energy enforces Eqs.~\eqref{eq:Nbar}--\eqref{eq:Ebar} even with
interaction~\cite{Taniguchi14}.  
Note, the expression of $\bar{N}$ may
be understood as a generalization of the Friedel sum
rule~\cite{Langer61,Langreth66,Rontani06} that holds at the zero
temperature.

Therefore the structure of Eqs.~\eqref{eq:LB1}--\eqref{eq:Ebar} is
intact even for the single-impurity Anderson model, in the
the wide-band limit. Accordingly, we can deduce that a small deviation
of $\Phi_{\text{ss}}$ should take a form of \eqref{eq:dPhi-ss}.
Equivalently, it can be written as
\begin{align}
& d \Phi_{\text{ss}} 
= \mp \int d\varepsilon\, \rho(\varepsilon)
  \sum_{a}\frac{\gamma_{a}}{\gamma}\, d \log \left[ 1 \mp e^{-\beta_{a}
  (\varepsilon - \mu_{a})} \right]
,\\ & \quad
= \int d\varepsilon\, \rho(\varepsilon)
  \sum_{a}\frac{\gamma_{a}}{\gamma} f_{a}(\varepsilon)\, d [
      \beta_{a} \mu_{a} - \beta_{a} \varepsilon].
\label{eq:dPhi-int}
\end{align}
The deviation $d[\beta_{a} \mu_{a} -\beta_{a} \varepsilon]$ is taken
by regarding 
$(\bar{\beta}, \overline{\beta\mu}, A^{N}_{a}, A^{E}_{a})$ 
as independent parameters, which gives Eq.~\eqref{eq:dPhi-ss}.  
The form \eqref{eq:dPhi-int} is surprising, when one recalls that the
local interaction makes the reduced density matrix
$\hat{\varrho}_{\text{red}}$ non-Gaussian, and the spectral function
$\rho(\varepsilon)$ dependent on the parameters. We suspect that there
is some cancellation between the quadratic and quartic contributions,
similarly to the nonequilibrium Ward identities~\cite{Oguri05},
because $\hat{\varrho}_{\text{ss}}$ still takes the Gaussian form
Eq.~\eqref{eq:rho-ss} in terms of scattering-state fields, even for an
interacting
dot~\cite{Hershfield93,Frohlich03,Tasaki06,Oguri07,Ness13,Ness14b}.
An actual mechanism is missing, though.

\subsection{Characterizing low-temperature heat current fluctuations}

In the context of the large deviation approach to equilibrium
statistical mechanics~\cite{EllisBook06}, the free energy
characterizes not only average quantities but also their fluctuations,
being the cumulant generating function (CGF).  Since the CGF of
steady-state currents has been
known~\cite{Levitov93,Levitov96,Saito07}, it would be desirable to see
a connection with $\Phi_{\text{ss}}$, but a generic link between the
CGF and $\Phi_{\text{ss}}$ is missing.  Nevertheless we can show that
the heat transport at low temperature give a concrete example of how
$\Phi_{\text{ss}}$ is capable of characterizing the fluctuations.  

The average heat current between two reservoirs (with different
temperatures $\beta_{1,2}$ and the same chemical potential
$\mu=0$) exhibits the universal behavior~\cite{Schwab00,Meschke06},
\begin{align}
& J^{Q} \approx \frac{c \pi}{12 \hbar} \left( \frac{1}{\beta_{1}^{2}} -
 \frac{1}{\beta_{2}^{2}} \right) ,
\end{align}
with the transmission coefficient $c=\mathcal{T}_{12}(\mu)$.  It is
known that the low-temperature current fluctuations are dominated by
the thermal (Johnson-Nyquist) noise, even for the extreme
nonequilibrium situation $\beta_{1}/\beta_{2} \gg 1$~\cite{Krive01}.
Since one can connect the thermal noise with the thermal conductance,
one may well say that $\Phi_{\text{ss}}$ characterizes those noises.

For perfect (or ``critical'') transmission $c\approx 1$ which is
realized for $\gamma_{1}$ = $\gamma_{2}$, one can develop conformal
field theory to construct the CGF $F(\lambda)$ to characterize
low-temperature heat current
fluctuations~\cite{Bernard12b,Bernard13,Bhaseen15}, which is found to
be
\begin{align}
& F(\lambda) \approx \frac{c\pi}{12 \hbar} \left[
  \frac{i\lambda}{\beta_{1}(\beta_{1}-i\lambda)} -
  \frac{i\lambda}{\beta_{2}(\beta_{2}+i\lambda)} \right].
\label{eq:F-lambda}
\end{align}
It corresponds to the the low-temperature limit of
Ref.~[\onlinecite{Saito07}].
What is interesting in the present context is that \citet{Bernard13}
have noticed that the function $dF(\lambda)/d\lambda$ is related with
the nonlinear heat current $J$ by what they call the extended
fluctuation relations. Then by comparing with Eq.~\eqref{eq:dPhi-ss},
we come to see that the CGF $F(\lambda)$ is directly given in terms of
$\Phi_{\text{ss}}$:
\begin{align}
F(\lambda) &= \frac{\gamma}{\hbar} \left[ \Phi_{\text{ss}}(\bar{\beta}, A^{E}+2i\lambda) - \Phi_{\text{ss}}(\bar{\beta}, A^{E}) \right].
\label{eq:F-by-Phi}
\end{align}
with $\bar{\beta} = (\beta_{1} + \beta_{2})/2$ and
$A^{E} = -\beta_{1}+ \beta_{2}$.  Indeed, the low-temperature behavior
of $\Phi_{\text{ss}}$ is readily evaluated from
Eqs.~\eqref{eq:Phi1-analytical} and \eqref{def:Phi-ss} as
\begin{align}
& \Phi_{\text{ss}} \approx \frac{c\pi}{12\gamma} \left(
  \frac{1}{\beta_{1}} + \frac{1}{\beta_{2}}\right), 
\end{align}
and putting it into Eq.~\eqref{eq:F-by-Phi} exactly reproduces
Eq.~\eqref{eq:F-lambda}. 
It is noted that the fluctuation theorem
$F(\lambda)=F(iA^{E}-\lambda)$ is equivalent to the inversion symmetry
$\Phi_{\text{ss}}(\bar{\beta},A^{E}) = \Phi_{\text{ss}}(\bar{\beta},
-A^{E})$ in this case. 

\section{Conclusion}

In summary, we have developed a thermodynamic description of the
nonequilibrium steady state that connects with multiple reservoirs,
and demonstrated that the steady-state Massieu-Planck function
$\Phi_{\text{ss}}$ can characterize consistently its quantum transport
properties of charge, energy or heat. The positive entropy production
rate caused by irreversible processes is also characterized by
$\Phi_{\text{ss}}$.  We have evaluated explicitly for a single-level
model that connects with multiple reservoirs, and argued that the same
thermodynamic structure persists even for a steady state with local
interaction, and that the heat current fluctuations are related to the
function $\Phi_{\text{ss}}$ at low temperature.

\begin{acknowledgments}
  The author gratefully acknowledges financial support from
  JSPS KAKENHI Grant Number 26400382.
\end{acknowledgments}

\appendix



\section{Thermodynamic relations of the equilibrium Massieu-Planck function}
\label{sec:MP-thermodynamics}

We recall some basic thermodynamic relations of the equilibrium
Massieu-Planck function $\Phi_{\text{eq}}$, which is defined as the
logarithm of the (grand) partition function~\cite{CallenBook85}.  With 
a given thermodynamic potential $\Omega(T,\mu)$ as a function of the
temperature $T$ and the chemical potential $\mu$, the function
$\Phi_{\text{eq}}$ can be written as
$\Phi_{\text{eq}} = -\Omega(T,\mu)/T$.  Below we show how it is
beneficial to regard $\Phi_{\text{eq}}$ as a function of the inverse
temperature $\beta=1/T$ and $\beta\mu$.  The volume of the system is
irrelevant and ignored because we treat a nanoscale system.

Starting with the thermodynamic relation $d\Omega = -N d\mu - S dT$,
the thermodynamic relation of $\Phi_{\text{eq}}$ becomes
\begin{align}
& d \Phi_{\text{eq}} =- \Omega d\beta + N \beta d\mu - S\frac{d\beta}{\beta} 
= N d(\beta \mu) - E d\beta,
\label{eq:dPhi-eq}
\end{align}
where we identify the entropy $S = -\beta \Omega + \beta
(E-N\mu)$. 
One can check the above directly by making a quantum statistical
construction of the partition function $e^{\Phi_{\text{eq}}}$.  For
noninteracting bosonic/fermionic particles with levels
$\{\epsilon_{\alpha}\}$, one finds
\begin{align}
& \Phi_{\text{eq}} (\beta,\beta\mu) = \mp \sum_{\alpha} \ln \left[ 1 \mp
  e^{-\beta(\epsilon_{\alpha} -\mu)} \right].  
\label{eq:Phi-isolated}
\end{align}
Then it is easy to see 
\begin{align}
\frac{\partial \Phi_{\text{eq}} (\beta,\beta\mu)}{\partial(\beta \mu)}
  \bigg|_{\beta} 
&= \sum_{\alpha} f(\epsilon_{\alpha}) = N, \\
 \frac{\partial \Phi_{\text{eq}} (\beta,\beta\mu)}{\partial \beta}
  \bigg|_{\beta\mu} 
    &= - \sum_{\alpha} \epsilon_{\alpha} f(\epsilon_{\alpha}) = -E, \\
\frac{\partial \Phi_{\text{eq}} (\beta, \beta\mu)}{\partial \beta} \bigg|_{\mu}
&
= -\left( E - \mu N \right),
\end{align}
where $f(\epsilon)=[e^{\beta(\epsilon-\mu)} \mp 1]^{-1}$ is the
distribution function.  The entropy
$S = -\partial \Omega/\partial T\big|_{\mu}$ is given by
\begin{align}
& S 
= \Phi_{\text{eq}} - \beta \frac{\partial \Phi_{\text{eq}}}{\partial
  \beta}\bigg|_{\mu} 
= \Phi_{\text{eq}} + \beta (E - \mu N).
\end{align}
The first equality means that $\Phi_{\text{eq}}$ is a Legendre
transform of $S$ regarding $\beta$ with a fixed $\mu$. One finds 
the entropy $S$ taking a familiar Shannon-like form: 
\begin{align}
& S = -\sum_{\alpha} \left\{f(\epsilon_{\alpha}) \ln
  f(\epsilon_{\alpha}) \mp \left[ 1\pm f(\epsilon_{\alpha})  \right]
  \ln \left[ 1\pm f(\epsilon_{\alpha})  \right] \right\}.
\label{eq:S-isolated}
\end{align}

\section{Effective single-level thermodynamics coupled with a single
  reservoir}
\label{sec:MP1-thermodynamics}

The calculation of the effective free energy of a single level
$\epsilon_{d}$ that couples with a single reservoir (with the inverse
temperature $\beta$ and the chemical potential $\mu$) has been long
known for fermionic systems~\cite{Schotte75} as well as bosonic
systems~\cite{WeissBook08,Hanggi08,Adamietz14}. As in equilibrium, we
then recast it to find the single-reservoir Massieu-Planck function
$\Phi_{1}$ in Eq.~\eqref{def:Phi1} of the main text.
Though the system coupled with a single reservoir no longer obeys the
Gibbs ensemble, we may regard the reservoir's parameters $\beta$ and
$\mu$ as local thermal parameters of this open system, because of the
thermodynamic relation, 
\begin{align}
& N = \frac{\partial \Phi_{1}(\beta,\beta\mu)}{\partial (\beta \mu)}
= \int^{\infty}_{-\infty} d\varepsilon \, \rho(\varepsilon)
  f(\varepsilon), 
\label{eq:N1}\\
& E = -\frac{\partial \Phi_{1} (\beta, \beta\mu)}{\partial \beta}
= \int^{\infty}_{-\infty} d\varepsilon \, \rho(\varepsilon)
  \varepsilon f(\varepsilon). 
\label{eq:E1}
\end{align}
Comparing between Eq.~\eqref{eq:Phi-isolated} and Eq.~(11) in the main
text, we see the spectral function $\rho(\varepsilon)$ play a role of
degeneracy at each energy shell. Physical quantities are expressed
simply by replacing the summation $\sum_{\alpha}( \cdots )$ by the
energy integral $\int d\varepsilon \rho(\varepsilon) (\cdots)$.
Particularly, the local entropy $S = \Phi_{1} + \beta (E-N\mu)$
becomes
\begin{align}
& S = -\int d\varepsilon \, \rho(\varepsilon)
  \left\{f(\varepsilon) \ln 
  f(\varepsilon) \mp \left[ 1\pm f(\varepsilon)  \right]
  \ln \left[ 1\pm f(\varepsilon)  \right] \right\}.
\end{align}
This form of the entropy indicates that some entities having distribution
$f(\varepsilon)$ is present at each energy shell, when putting it in
the context of the information theory.   This is why we have introduced
field operators at each energy shell in the main text.

\section{Analytical form of $\Phi_{1} (\beta,\beta\mu)$ and
  regularization}
\label{sec:MP1-analytical}

For the Lorentzian spectrum of Eq.~\eqref{def:spectral-fn} in the main text, 
it is possible to obtain the analytical form of the single-reservoir
Massieu-Planck function $\Phi_{1}$, hence the steady-state
$\Phi_{\text{ss}}$ via Eq.~\eqref{def:Phi-ss} in the main text.  
The function $\Phi_{1}$ turns out to be divergent due to the
zero-temperature contribution in the wide-band limit.~\footnote{As for
  bosonic transport, we don't usually need such regularization because
  we can set $\mu=0$ with the $\varepsilon$-integral spanning
  $(0,+\infty)$. Accordingly all the finite contributions are halved for that
  case.}  To suppress such divergence, we need to introduce a finite
band width of the reservoir.  The explicit form of $\Phi_{1}$ is
useful to connect several different expressions found in the
literature; it also clarifies the nature of the divergence and shows
directly that physical quantities are independent of the
regularization of such divergence.

A central role is played by the following integral formula
\begin{align}
& \int^{\infty}_{-\infty} \frac{d\varepsilon}{(\varepsilon - \epsilon_{d} +
  i\gamma)(e^{\beta(\varepsilon - \mu)} \mp 1)}
\notag \\ & \quad 
= \int^{\infty}_{-\infty} \frac{\mp \theta(\mu - \varepsilon) \, d\varepsilon}{\varepsilon - \epsilon_{d}
  + i\gamma}- \frac{d}{dz}\log \mathcal{F}(z)
,
\label{eq:the-integral}
\end{align}
where the first term corresponds to the zero-temperature contribution
while the second term, to the finite-temperature. The latter can be
evaluated explicitly (see \citet{GradshteynBook14} 3.415). We find it useful
to express it by the complex function $\mathcal{F}(z)$ that is normalized
by $\mathcal{F}(z) \to 1$ for large $|z|$:
\begin{align}
& \mathcal{F}(z) =
  \begin{cases}
\displaystyle{\frac{\Gamma(z) \,
  z^{\frac{1}{2}-z} e^{z}}{\sqrt{2\pi}}}, & \text{(B)}, \\
\displaystyle{\frac{\sqrt{2\pi} z^{z} e^{-z}}{\Gamma(\frac{1}{2}
  + z)}}, & \text{(F)},
  \end{cases}
\label{def:F-z}
\end{align}
as a function of the dimensionless complex parameter
\begin{align}
& z= x + iy = \frac{\beta}{2\pi} (\gamma + i\xi); \quad \xi =
  \epsilon_{d} - \mu.
\label{def:z}
\end{align}
The first term on the right-hand side of Eq.~\eqref{eq:the-integral}
is divergent, which we need to suppress  by introduce finite band width of the
reservoir,
\begin{align}
& \int^{\infty}_{-D} \frac{\mp \theta(\mu - \varepsilon) \,
  d\varepsilon}{\varepsilon - \epsilon_{d} + 
  i\gamma} = \mp \log \big(z/z_{D} \big),
\end{align}
with the dimensionless cutoff $z_{D} = \beta [\gamma +
i(D+\epsilon_{d})]/2\pi$. 
The formula \eqref{eq:the-integral} allows
us to evaluate the average occupation number $N$ and energy $E$ in
Eqs.~\eqref{eq:N1}--\eqref{eq:E1} straightforwardly:
\begin{align}
& N(z) = \frac{1}{\pi} \Im \left[ \pm \log \big( z/z_{D} \big) +
  \frac{d}{dz} \log 
  \mathcal{F}(z) \right]
\label{eq:N-by-F} 
,\\ & \quad
= \begin{cases}
-\frac{1}{2} + \frac{1}{\pi} \Im \left[\psi(z) +  \frac{1}{2z} \right],
& \text{(B)}, \\
\frac{1}{2} - \frac{1}{\pi} \Im \left[ \psi(\frac{1}{2} + z)
\right],  & \text{(F)},
\end{cases}
\label{eq:N-by-psi} \\ 
& E(z) = \frac{1}{\pi} \Im \left[ (\epsilon_{d}-i\gamma) \left( \pm \log
  \big( z/z_{D} \big)   +
  \frac{d}{dz} \log 
  \mathcal{F}(z)  \right)\right]
\label{eq:E-by-F} 
,\\ & \quad
= \begin{cases}
\frac{1}{\pi} \Im \left\{(\epsilon_{d} - i\gamma) \left[
  \psi(z) + \frac{1}{2z} - \log z_{D} \right] \right\}
,& \text{(B)}, \\
-\frac{1}{\pi} \Im \left\{(\epsilon_{d} - i\gamma) \left[
  \psi(\tfrac{1}{2} + z) - \log z_{D} \right] \right\}
,  & \text{(F)}.
  \end{cases}
\label{eq:E-by-psi}
\end{align}
It is noted that while the average energy $E$ reduces to
$\epsilon_{d} f(\epsilon_{d})$ in the isolated limit $\gamma\to +0$, it
diverges for any finite $\gamma$ because of the zero-temperature
contribution. Yet its finite-temperature contribution is well-defined.

One can construct the function $\Phi_{1}$ by integrating the above
expression of $N$ or $E$. Those forms suggest that we may write it as
\begin{align}
& e^{\Phi_{1}} = A\; |\mathcal{F}(z) |^{2},
\end{align}
where $A$ is a factor coming from the zero-temperature contribution.
Seeing $\Phi_{1}$ vanish at large $y$, one determines $A$ as
\begin{align}
  & \log A = \pm 2 \Re \int^{z_{D}}_{z}\!\! dz' \log (z'/z_{D})
= \left| \left(z_{D}/z\right)^{z}\right|^{\mp 2}.
\end{align}
This leads to the single-reservoir Massieu-Planck function,
\begin{align}
& \Phi_{1} (\beta, \beta\mu) = 
\begin{cases}
\displaystyle{\ln \left| \frac{z^{\frac{1}{2}}\, 
      \Gamma(z)}{\sqrt{2\pi}  z_{D}^{z}\, 
  e^{-z}} \right|^{2}}, & \text{(B)}, \\
\displaystyle{ \ln \left| \frac{\sqrt{2\pi} z_{D}^{z}\, 
e^{-z}}{\Gamma(\tfrac{1}{2} + z)} \right|^{2}}, & \text{(F)}.
  \end{cases}
\label{eq:Phi1-analytical}
\end{align}

It is always good to check the isolated limit $\gamma \to +0$. In this
limit, we find
\begin{align}
\left| \mathcal{F}(x + iy)  \right|^{2} &\:\to\:
e^{\pm \pi(|y|-y)} \left( 1 \mp e^{-2\pi y} \right)^{\mp 1}, \\
A \quad &\:\to\:  e^{\pm \pi(y-|y|)}.
\end{align}
Hence $e^{\Phi_{1}} \to \left[ 1 \mp e^{-2\pi y} \right]^{\mp
  1}$. This is nothing but the partition function of the isolated
system Eq.~\eqref{eq:Phi-isolated}.

In addition to $N$ and $E$, we can obtain various thermodynamic
quantities by differentiating $\Phi_{1} (\beta,\beta\mu)$. They take a
simple form in terms of the function $\mathcal{F}(z)$. For instance,
the entropy $S$ becomes
\begin{align}
S &= \Phi_{1} -\beta \frac{\partial \Phi_{1}}{\partial \beta} \bigg|_{\mu}
= 2\Re \bigg[\Big( 1-z \frac{d}{dz} \Big) \log \mathcal{F}(z) \bigg]
, \\ &
=  \begin{cases}
2 \Re \left[ \log \left( \frac{\Gamma(z)
     z^{\frac{1}{2}}}{\sqrt{2\pi}}  \right) + z  -\frac{1}{2} - z \psi
     (z) \right], & \text{(B)} \\
2 \Re \left[ \log \left( \frac{\sqrt{2\pi}}{\Gamma(\frac{1}{2}+z)}
     \right) - z  + z \psi (\tfrac{1}{2}+z) \right]. & \text{(F)}
  \end{cases}
\end{align}
The vanishing of $S$ at zero temperature follows from the fact
$\mathcal{F}(z) \to 1$ for large $z$. 
The specific heat $C$ becomes
\begin{align}
C &= \beta^{2} \frac{\partial^{2} \Phi_{1}}{\partial
  \beta^{2}}\Big|_{\mu} 
 = 2 \Re \left[ z^{2} \frac{d^{2}}{dz^{2}}\log \mathcal{F}(z)  \right]
, \\ &
=  \begin{cases}
2 \Re \left[ -\frac{1}{2} - z + z^{2} \psi'(z) \right] & \text{(B)} \\
2 \Re \left[ z - z^{2} \psi'(\tfrac{1}{2} + z) \right] & \text{(F)}
\end{cases}.
\end{align}
The result of specific heat for fermionic systems agrees with
that of \onlinecite{Schotte75} (with $\xi=0$), while
for the bosonic systems, it agrees with that
of the damped harmonic oscillator~\cite{Hanggi08,Adamietz14}.

\section{Nonlinear current of charge, energy and heat}
\label{sec:currents-analytical}

One can also find analytical expressions of nonlinear currents of
particle, energy, or heat; they are well-defined and
independent of the cutoff.  From the thermodynamic
relation~\eqref{eq:dPhi-ss} [or equivalently from
Eqs.~\eqref{eq:LB1}--\eqref{eq:LB2}] in the main text, we can write currents of particle and
energy as 
\begin{align}
& I_{a} = \frac{2}{\hbar} \sum_{b}
  \frac{\gamma_{a} \gamma_{b}}{\gamma}
\left[ N(z_{a}) - N(z_{b}) \right]
\label{eq:I-analytical}, \\
& J_{a} = \frac{2}{\hbar} \sum_{b}
  \frac{\gamma_{a} \gamma_{b}}{\gamma}
 \left[ E(z_{a}) - E(z_{b}) \right].
\label{eq:J-analytical}
\end{align}
Here $N(z_{a})$ and $E(z_{a})$ are defined by
Eqs.~\eqref{eq:N-by-F}--\eqref{eq:E-by-psi} for each reservoir by
choosing
$z_{a} = (\beta_{a}/2\pi) (\gamma + i\epsilon_{d} - i\mu_{a})$ with
$\beta_{a}$ and $\mu_{a}$ of the reservoir [see Eq.~\eqref{def:z}].
The difference of $N$ or $E$ is independent of cutoff.  We see
\begin{align}
& N(z_{a})-N(z_{b}) = \frac{1}{\pi} \Im \left[ \pm \log
  \Big(\frac{z_{a}}{z_{b}}　\Big) + 
  \frac{\mathcal{F}'(z)}{\mathcal{F}(z)} \bigg|^{z_{a}}_{z_{b}} \right],
\\
& E(z_{a}) - E(z_{b}) 
\notag \\ & \quad
= \frac{1}{\pi} \Im \left\{
  (\epsilon_{d}-i\gamma) \left[ \pm \log
  \Big(\frac{\beta_{b}z_{a}}{\beta_{a} z_{b}} \Big) +
  \frac{\mathcal{F}'(z)}{\mathcal{F}(z)} \bigg|^{z_{a}}_{z_{b}}
\right]\right\}.
\end{align}
Therefore the finite-temperature contribution are written in terms of the
digamma function $\psi(z) = \Gamma'(z)/\Gamma(z)$ via Eq.~(\ref{def:F-z}).  Likewise, heat current $J_{a}^{Q}=J_{a} -
\mu_{a} I_{a}$ from the reservoir $a$ can be expressed analytically by
using the above expressions. 

\section{Derivation of Eq.~\eqref{eq:NE-by-Phi} or
  \eqref{eq:currents-by-Phi}} 
\label{sec:some-detail}

Assigning an affinity that is associated with each current turns out
delicate, particularly in multiterminal setting. Because of the
current conservation, the inflow at the reservoir $a$ must involve
outflows to other reservoirs.  In order to establish
Eqs.~\eqref{eq:NE-by-Phi}, \eqref{eq:currents-by-Phi}, it
is crucial to specify how one varies a relevant parameter by fixing
others, as in equilibrium thermodynamics.

In the following, we choose to use $x_{a}$ for $\beta_{a} \mu_{a}$ or
$-\beta_{a}$, to describe particle or energy transport, while we will
introduce $y_{a}$ for affinities later.
Adopting this notation, we write the steady-state Massieu-Planck function
as
$\Phi_{\text{ss}} = \sum_{a} (\gamma_{a}/\gamma) \Phi_{1} (x_{a})$,
where $\Phi_{1}$ is defined by Eq.~\eqref{def:Phi1} in the text.
Now we define the local parameter $\bar{x}$ and its affinity $y_{a}$
by
\begin{align}
& \bar{x} = \sum_{a=1}^{N} \frac{\gamma_{a}}{\gamma} x_{a}; \quad
y_{a} = x_{a} - \bar{x}.
\end{align}
Affinities $y_{a}$ satisfy the sum rule
$\sum_{a} (\gamma_{a}/\gamma) y_{a} = 0$. This comes from the
condition that any variation of affinities does not affect $\bar{x}$.
In this way, we map the $N$ parameters $\{x_{a}\}$ into $\bar{x}$ and
$\{ y_{a} \}$, where we can eliminate one of
$\{y_{a}\}$. Alternatively, we can express $x_{a}$ as a function of
$\bar{x}$ and the differences of $\{y_{a}\}$ by
\begin{align}
& x_{a} = \bar{x} + \sum_{b=1}^{N} \left( \delta_{ab} - \frac{\gamma_{b}}{\gamma} \right) y_{b} 
= \bar{x} + \sum_{b=1}^{N} \frac{\gamma_{b}}{\gamma} (y_{a} - y_{b}).
\label{eq:x-and-y}
\end{align}
This parametrization is singled out by requiring to fix $\bar{x}$ by
any variation of the $N$-independent parameters
$\{y_{1}, \ldots y_{N}\}$.  Though we can safely cross out one of
$\{y_{a}\}$ at any moment, we prefer retaining all of them for a
symmetrical reason.  In either way, we can obtain the currents
fulfilling the current conservation by varying $y_{a}$ in the above. 

Equation~(22) or (23) can be derived by taking derivatives
regarding $\bar{x}$ or $y_{a}$ by assuming $\bar{x}$ and $y_{a}$ are
independent variables:
\begin{align}
& \bar{\mathcal{N}} = \frac{\partial \Phi_{\text{ss}}}{\partial
                \bar{x}}
= \sum_{a=1}^{N} \frac{\gamma_{a}}{\gamma} \frac{\partial x_{a}}{\partial
                \bar{x}} \frac{\partial \Phi_{1}(x_{a})}{\partial
                x_{a}}
= \sum_{a=1}^{N} \frac{\gamma_{a}}{\gamma} \mathcal{N}_{a}, \\
& \mathcal{I}_{a} = \frac{2}{\hbar} \frac{\partial \Phi_{\text{ss}}}{\partial
                                                           y_{a}}
= \frac{2\gamma}{\hbar} \sum_{b=1}^{N} \frac{\gamma_{b}}{\gamma} 
\frac{\partial x_{b}}{\partial y_{a}} \frac{\partial \Phi_{1}(x_{b})}{\partial x_{b}}
, \\ & \quad
= \frac{2\gamma}{\hbar} \sum_{b=1}^{N} \frac{\gamma_{b}}{\gamma} 
\left( \delta_{ab} - \frac{\gamma_{a}}{\gamma} \right) \mathcal{N}_{b}
= \frac{2}{\hbar} \sum_{b=1}^{N} \frac{\gamma_{a}\gamma_{b}}{\gamma} \left( \mathcal{N}_{a} - \mathcal{N}_{b} \right),
\label{eq:current-form}
\end{align}
Here we assign, for particle transport, $\bar{\mathcal{N}}=\bar{N}$,
$\mathcal{I}_{a}=I_{a}$, and $\mathcal{N}_{a}=N(z_{a})$; for energy
transport, $\bar{\mathcal{N}}=\bar{E}$, $\mathcal{I}_{a}=J_{a}$, and
$\mathcal{N}_{a}=E(z_{a})$.  The current conservation
$\sum_{a} \mathcal{I}_{a} = 0$ immediately follows from the above
expression of $\mathcal{I}_{a}$.  

Linear current $\mathcal{I}_{a}$ at terminal $a$ is proportional to
$\gamma_{a} (x_{a} - \bar{x}) = \gamma_{a} y_{a}$. Choosing the
parametrization~\eqref{eq:x-and-y}, it shows that positive affinity
$y_{a}$ plays a role of inducing the inflow at the terminal $a$ but
the outflows at all the other terminals, in the linear-response
regime.

\bibliographystyle{apsrev} 

\end{document}